\documentclass[a4paper, amsfonts, amssymb, amsmath, reprint, showkeys, nofootinbib, twoside]{revtex4-2}
\usepackage[english]{babel}
\usepackage[utf8]{inputenc}
\usepackage{xcolor}

\usepackage{graphicx}
\graphicspath{{figures/}}

\usepackage[squaren]{SIunits}
\usepackage{tabularx}
\usepackage{lipsum}

\newcommand{\timescoupling}{$\cdot 10^{16}\,\textrm{W}\textrm{K}^{-1}\textrm{m}^{-3}$}
\renewcommand{\sp}{\textit{sp}}
\newcommand{\del}{\textit{d}}
\newcommand{\deltan}{$\Delta n$}

\renewcommand{\one}{(i)}
\newcommand{\two}{(ii)}
\newcommand{\three}{(iii)}

\begin{document}

\title{Influence of band occupation on electron-phonon coupling in gold}
\author{Tobias Held}
\author{Sebastian T. Weber}
\author{Baerbel Rethfeld}
\affiliation{Department of Physics and OPTIMAS Research Center, RPTU Kaiserslautern-Landau, Gottlieb-Daimler-Stra\ss e 76, 67663 Kaiserslautern, Germany}

\date{August 1, 2023}

\begin{abstract}
    Electron-phonon coupling is a fundamental process that governs the energy relaxation dynamics of solids excited by ultrafast laser pulses.
    It has been found to strongly depend on electron temperature as well as on nonequilibrium effects.
    Recently, the effect of occupational nonequilibrium in noble metals, which outlasts the fully kinetic stage, has come into increased focus.
    In this work, we investigate the influence of nonequilibrium density distributions in gold on the electron-phonon coupling.
    We find a large effect on the coupling parameter which describes the energy exchange between the two subsystems.
    Our results challenge the conventional view that electron temperature alone is a sufficient predictor of electron-phonon coupling.
\end{abstract}

\maketitle

\section{\label{sec:intro}Introduction}
Solid-state physics has been revolutionized by the development of ultrashort laser pulses, which make it possible to manipulate and probe matter on unprecedented timescales \cite{Strickland1985, Fann1992b, Krausz2009, Bauer2015}.
This technological breakthrough has given rise to new fields of research, such as ultrafast magnetization switching and transient states of matter \cite{Beaurepaire1996,Stanciu2007,Koopmans2010,Sokolowski-Tinten1998, Lindenberg2005, Chapman2011, Mo2018}.
In industrial processes, ultrashort pulses are essential for laser ablation and precision material processing \cite{BaeuerleBuch11,Balling2013,Ivanov2015,Shugaev2016}.

To fully exploit these new technologies, a thorough understanding of the underlying mechanisms is essential.
An optical femtosecond-laser irradiating the surface of a solid directly excites only the electron system of the material.
The subsequent energy transfer to the phonons occurs on a picosecond timescale.
The most widely used model to describe excitation and relaxation is the rather simple Two-Temperature Model (TTM) \cite{Anisimov1974}, which has also inspired several extensions \cite{Beaurepaire1996, vanDriel1987,  Carpene2006, Mueller2014PRB, Waldecker2016}.
Its central parameter is the material-specific electron-phonon coupling parameter, which determines the rate of energy transfer and thus the relaxation dynamics of the solid.
In literature, the electron-phonon coupling has been found to depend on many material parameters \cite{Medvedev2020}, with a focus on the electron temperature-dependence \cite{Wang1994,Lin2007,Petrov2013,Brown2016,Smirnov2020}.

Our previous calculations using Boltzmann collision terms have shown that the electron-phonon coupling is strongly influenced by nonthermal effects, which persist on a sub-100\,fs timescale after the laser irradiation~\cite{Rethfeld2002, Mueller2013PRB, Mueller2014ASS, Weber2017}.
Following this fully kinetic stage, the electrons are generally Fermi-distributed~\cite{Fann1992b,Rethfeld2002}.
However, the occupation of the bands can still be out of equilibrium~\cite{vanDriel1987, Murray2007, Raemer2014, Ndione2022}, which can significantly alter the electronic properties.
Recent results have indicated that such occupational nonequilibria play an important role in describing the transient optical properties of noble metals~\cite{Ndione2022}.
In this work we extend our previous model \cite{Mueller2013PRB,Mueller2014ASS,Weber2017} to examine the influence of nonequilibrium band occupation on the electron-phonon coupling parameter.
The results are presented for gold, revealing a large influence of the occupational nonequilibrium on the electron-phonon coupling strength.

\section{\label{sec:model}The Multi-Band Model for Electron-Phonon Interaction}

We consider a solid with multiple electron and phonon bands, which is irradiated by an ultrashort laser pulse in the visible spectrum.

\subsection{Coupling of Multiple Bands}

The model equations involve a number of separate electron bands and one phonon band, each described by a temperature.
Analogously to the two-temperature model (TTM) \cite{Anisimov1974}, the electrons' temperatures are coupled to the phonons
\begin{align}
\frac{d{u_{e}^{i}}}{d{t}}\Bigr|_{\textrm{el-ph}} &= G_{ep}^{i} (T_e^i -T_p) \label{model_eq_el} \\
\frac{d{u_{p}}}{d{t}}\Bigr|_{\textrm{el-ph}} &= \sum_{i} G_{ep}^{i} (T_p -T_e^i) \label{model_eq_ph},
\end{align}
where we introduce the partial electron-phonon coupling parameters $G_{ep}^{i}$ for the individual electron band $i$.
Here, $u$ denotes a subsystem's internal energy and $T$ its temperature; index $e$ refers to electrons and $p$ to phonons.
Note that a similar model has been proposed in Ref.~\cite{Waldecker2016}, which, however, considers several phonon modes with respective temperatures.

If the energy transfer between an electron and phonon band is known, the corresponding partial electron-phonon coupling parameter $G_{ep}^{i}$ can be obtained by solving Eq.~(\ref{model_eq_el}) for it,
\begin{align}
    G_{ep}^{i} &= \frac{1}{(T_{p} -T_{e}^{i})} \frac{d{u_{e}^i}}{d{t}}\Bigr|_{\textrm{el-ph}}
	\label{partial_gep} .
\end{align}

\subsection{How to calculate the coupling}

The energy change of band $i$ due to electron-phonon interaction can be computed by integrating the temporal derivative of the distribution function

\begin{equation}
	\frac{d{u_{e}^i}}{d{t}}\Bigr|_{\textrm{el-ph}} =
	\int d{E}\, \frac{d{f^{i}}}{d{t}}\Bigr|_{\textrm{el-ph}}D^{i}(E)E \enspace ,
	\label{u_change}
\end{equation}
with the partial density of states (DOS) $D^{i}(E)$, which is assumed to be constant in time.

In this work, we obtain this derivative of the distribution function for each band from a Boltzmann collision integral

\begin{equation}
    \frac{\partial{f^{i}}}{\partial{t}}\Bigr|_{\textrm{el-ph}} =
	\frac{2\Omega\pi^{3}}{\hbar k}\int d{E_{q}}\, |M_{ep}(q)|^{2} \frac{D_{ph}(E_{q})}{q} \mathcal{F}^{\pm} \Xi^{\pm} \frac{D^{i}(E^\pm)}{k^\pm} \enspace .
    \label{dfdtPW}
\end{equation}
Here, $q$ is the phonon momentum and $E_q$ the corresponding energy, $\Omega$ is the volume of the unit cell, $\mathcal{F}$ is the collision functional which ensures Pauli's principle, $\Xi$ represents the momentum conservation and $M_{ep}$ is the electron-phonon matrix element.
For the latter, we use an analytical screened plane-wave approach
\begin{equation}
    |M_{ep}(q)|^{2} = \frac{e^{2}}{2\epsilon_{0} \Omega} \frac{E_{q}}{q^{2}+\kappa^{2}} \enspace .
    \label{PWmatrix}
\end{equation}
with the screening parameter $\kappa$.
The screening has a considerable influence on the electron-phonon coupling strength, as will be discussed below.
This matrix element is derived for longitudinal acoustic phonons as it vanishes for the transversal modes.
For more details on this approach, see Ref.~\cite{Mueller2013PRB}.

For the determination of the coupling parameter, we only consider electronic intraband transitions.
Despite this limitation, the bands are not independent since they each contribute to the screening felt by all electrons.

\section{Application to Gold}

We apply the Multi-Band Model to Gold.
The following section contains all the information necessary to describe the material in the framework of our model.

\subsection{Types of electrons}

In gold, there are two types of electrons with different properties excitable by visible light:
Delocalized electrons resembling \sp{}-orbital states and bound electrons rather corresponding to \del{}-orbital states.
The flat $E(k)$ dispersion of the \del{}-band leads to sharp peaks in the density of states(DOS) of the \del{}-band while the \sp{}-band is closer to the DOS of a free electron gas .

\subsection{The Bandstructure}

There is no unique distinction between the two electron types.
Here, we used a DOS calculated with density functional theory (DFT) using Elk \cite{elk} with a projection onto \sp{}- and \del{}-states \cite{Ndione2019, Ndione2022}.
This leads to the \del{}-band DOS extending beyond the Fermi energy and the electrons not being completely localized at low temperatures.
Further discussion on the appropriate distinction between electron types might be required.
\begin{figure}[htbp]
    \centering
    \includegraphics[]{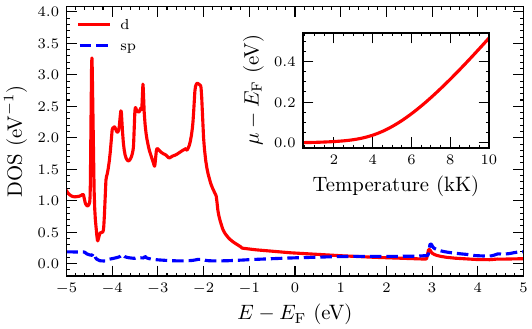}
    \caption{Band-resolved DOS of gold calculated with DFT using Elk \cite{elk}, the same as used in \cite{Ndione2019, Ndione2022}. The inset shows the equilibrium chemical potential, found with Eq.~(\ref{mu}).
    }
    \label{au_dos}
\end{figure}
Fig.~\ref{au_dos} shows the band-resolved DOS of gold used in this work.
The \del{}-band peaks reach up to about 2 eV below the Fermi energy, so the numerous electrons located in that region are immobile in equilibrium at low temperatures.

The dispersion relations $k^i(E)$ corresponding to each band $i \in {\sp{}, \del{}}$ are calculated from the partial DOS $D^i(E)$ using the effective one-band model \cite{Mueller2013PRB},
\begin{equation}
	k^i(E) = \sqrt[3]{6 \pi \int_0^E d{\epsilon} D^i(\epsilon)} \enspace .
	\label{effective_one_band}
\end{equation}
Compared to previous calculations for gold using a single band, this approach yields lower wavenumbers for both bands.
Since the one-band model tends to overestimate the wavenumbers, this should represent an improvement towards a more realistic description.
The dispersion enters the collision integral (\ref{dfdtPW}) and thus directly influences the resulting coupling parameter.

\subsection{Densities}

The density distribution between the bands generally depends on the temperature and the chemical potential.
In equilibrium, all electrons in a material follow a Fermi distribution, given by 
\begin{equation}
       f(E, T) = \left[e^{\frac{E-\mu_{eq}(T)}{k_B T}}+1\right]^{-1}
    \label{fermi_distribution}
\end{equation}
with a shared temperature $T$ and the temperature-dependent equilibrium chemical potential $\mu_{eq}(T)$, which is implicitly determined by the conservation of the total electron number $n$,
\begin{equation}
    n = \int d{E}\, f(E, T_e, \mu_{\mathrm{eq}}(T_e))\, D(E) = \textrm{const} \enspace ,
    \label{mu}
\end{equation}
with the total electron DOS $D(E) = \sum_{i} D^{i}(E)$.
Eq.~(\ref{mu}) can be numerically solved for $\mu_{eq}$ with a root-finding algorithm.
The resulting equilibrium chemical potential for the given DOS of gold is displayed as an inset in Fig~\ref{au_dos}.

The equilibrium chemical potential allows the calculation of the partial equilibrium densities $n_{\mathrm{eq}}^i(T_e)$ at each temperature,
\begin{equation}
    n_{\mathrm{eq}}^i(T_e) = \int d{E}\, f(E, T_e, \mu_{\mathrm{eq}}(T_e)) D^{i}(E) \enspace .
    \label{partial_n}
\end{equation}
Under laser excitation with visible light, an occupational nonequilibrium can be induced.
This means that the partial densities deviate from $n_{\mathrm{eq}}^i(T_e)$ and the bands do not share the same chemical potential \cite{Ndione2022}.
In most cases for excitation with visible light, this nonequilibrium excitation involves a depopulation of \del{}-band states, leading to on average more \sp{}-electrons than in the equilibrium case.
Here, we additionally consider a overpopulation of the \del{}-band, which might be reached in more complex excitation conditions.

\subsection{Phonons}

The phonons are assumed to be fixed at 300\,K.
They are described by a Debye model, using the same parameters as Ref. \cite{Mueller2013PRB} to facilitate the comparison.

\section{Results for Gold}

As a first test of the model, we consider occupational equilibrium, so the case where both bands share the same chemical potential. 
This allows a direct comparison between the presented two-band model and the description of gold through one band only.

\begin{figure}[tbhp]
    \centering
    \includegraphics[]{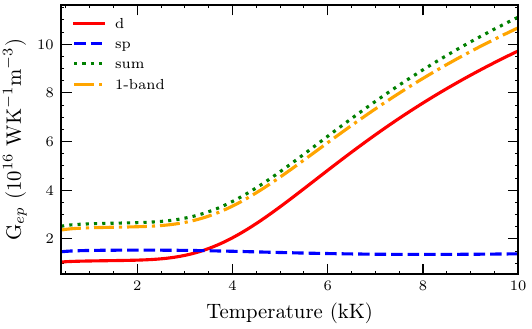}
    \caption{
    Total electron-phonon coupling of gold at occupational equilibrium in dependence on electron temperature.
    We see the comparison between the result with a 1-band approach, the individual results for \del{}- and \sp{}-band and their sum.
    The sum matches the 1-band result very well.
    }
    \label{au_eq}
\end{figure}

Fig.~\ref{au_eq} displays the coupling parameter of gold for the equilibrium density in dependence on electron temperature.
The yellow line shows the result for gold as a single electron band, the red and blue lines for the partial bands individually and the green line is the sum of the partial coupling parameters.

The total coupling is constant below 3000\,K and increases monotonously at higher temperatures, reaching up to 11 \timescoupling{}.
This behavior can be explained as a result of the Fermi edge broadening with temperature, which allows more electrons in the \del{}-band DOS peaks to contribute to the coupling.
This is confirmed by the observation that the \sp{}-coupling stays mostly constant over temperature,
while the \del{}-electrons are solely responsible for the increase of the coupling with temperature.
Overall, we see that the description of the coupling with two bands exclusively through intraband collisions reproduces the coupling of the equilibrium case very well.

Next, occupational nonequilibria are considered in addition to the temperature dependence.
We use the difference of the \del{}-band density to its 0\,K value, \deltan{}
\begin{align}
    \Delta n &= n^{d} - n^{d}_{eq}(0\,K) \\
    &= n^{sp}_{eq}(0\,K) - n^{sp} \enspace ,
    \label{deltan}
\end{align}
to describe the density distribution between the bands.
This is sufficient to describe the system, because the total particle number is conserved.
Positive \deltan{} represents high occupation of the \del{}-band and low occupation of the \sp{}-band and vice versa.

For the ease of visualization, we will only present the case where all electron bands share the same temperature.
In this case, a total coupling parameter can be obtained as
\begin{equation}
    G_{ep}^{\textrm{total}} = \sum_{i} G_{ep}^{i}
    \label{gep_sum} .
\end{equation}

\begin{figure}[tbhp]
    \centering
    \includegraphics[]{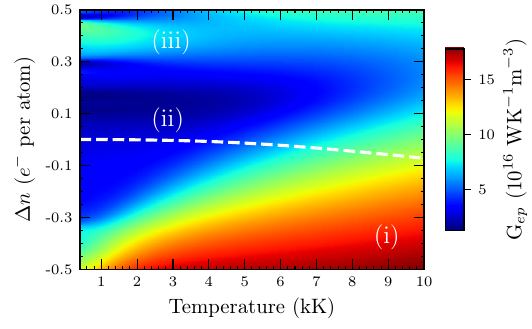}
    \caption{
    Color-coded total electron phonon coupling of gold in dependence on electron temperature and density distribution (described by Eq.~(\ref{deltan})).
    It is obtained from the sum of the individual coupling parameters.
    The dashed white line indicates the equilibrium density, for which the coupling is shown in Fig.~\ref{au_eq}.
    The roman numerals label regions of different coupling behavior and will be used as reference in the analysis below.
    }
    \label{au_sum}
\end{figure}

Fig.~\ref{au_sum} shows the total coupling parameter of gold, obtained by summing up the contributions from both bands, Eq.~(\ref{gep_sum}).
The coupling is color-coded and shown in dependence on electron temperature and the density distribution between the bands described by Eq.~(\ref{deltan}).
The white dashed line in Fig.~\ref{au_sum} indicates occupational equilibrium, for which the coupling has been analyzed in Fig.~\ref{au_eq}.
In this color plot, three density regions of distinctly different behavior emerge, which have been labeled by roman numerals.
In the region of low \deltan{}, meaning low \del{}-density and marked as \one{} in Fig.~\ref{au_sum}, the coupling increases monotonously with increasing temperature and with decreasing \del{}-density.
The temperature dependence is less pronounced at lower densities and higher temperatures.
Within this region, the coupling parameter reaches values of 16 \timescoupling{} and above, which is 50\% larger than the maximum value in equilibrium (Fig.~\ref{au_eq}).
In the central region, marked as \two{}, the coupling increases monotonously with temperature while maintaining a relatively low magnitude.
The density dependence in this region is negligible.
In region \three{}, where \deltan{} is the highest, the behavior of the coupling parameter is more complex.
The temperature dependence in this region varies with the precise density, showing both increasing and decreasing trends.
We find both a local maximum and minimum which are washed out at higher temperatures.
In terms of magnitude, the coupling covers values within the range seen in the equilibrium case (see Fig.~\ref{au_eq}), neither reaching its maximum nor minimum.

\begin{figure}[tbhp]
    \centering
    \includegraphics[]{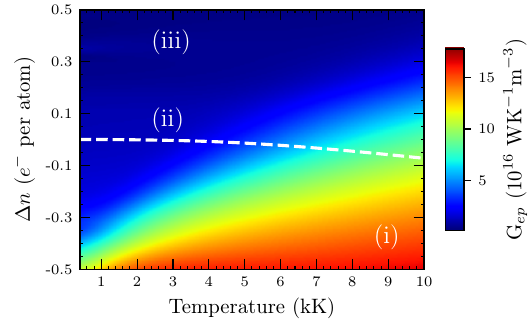}
    \vskip0.4cm
    \includegraphics[]{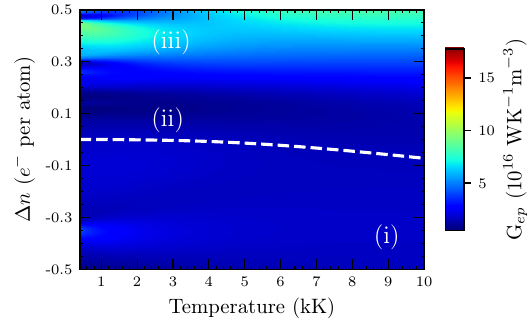}
    \caption
    {
    Partial electron-phonon coupling parameters of gold in dependence on temperature and density.
    Top: coupling of the \del{}-band electrons. Bottom: coupling of the \sp{}-band electrons.
    The coupling of the \del{}-electrons increases monotonously with increasing temperature and with decreasing density.
    The coupling of the \sp{}-electrons shows a complex non-monotonous behaviour.
    The additional labeling in both panels is the same as in Fig.~\ref{au_sum}.
    }
    \label{au_partial}
\end{figure}

To understand the behavior of the coupling, we take a closer look at the individual contributions from the bands.
Fig.~\ref{au_partial} (top) shows the coupling parameter of the \del{}-electrons plotted over electron temperature and density distribution, see Eq.~(\ref{deltan}).
For low \del{}-band densities (corresponding to area \one{}), the \del{}-band coupling parameter increases with temperature and strongly increases with decreasing \del{}-density.
The underlying mechanism for this behavior is an increased phase space for \del{}-electrons to scatter into.
At any given temperature for a subsystem, a lower density corresponds to a lower chemical potential.
Thus, a lower \del{}-density brings the Fermi edge closer to the DOS peaks, allowing a much larger amount of electrons to contribute to the coupling via collisions.
The behavior of the \del{}-band coupling in this region \one{} strongly resembles the one seen from the total coupling in Fig.~\ref{au_sum}.
The \del{}-band appears to dominate the increase in total coupling in this density region, contributing up to 90\% of the coupling.
However, in the regions of higher \del{}-occupation (areas \two{} and \three{}), the \del{}-band coupling almost vanishes with a negligible temperature dependence.
Note that the magnitude of the \del{}-band coupling could generally be overestimated in our approach:
The \del{}-electron wave functions differ significantly from plane waves, for which the applied matrix element was derived.
A lower coupling between \del{}-electrons and phonons as compared to \sp{}-electrons has also been determined in ab-initio calculations \cite{Brown2016}. 
Further indications for a low electron-phonon coupling strength in gold have been found experimentally \cite{White2014, Pudell2018, Mo2018}. 

Figure \ref{au_partial} (bottom) shows the electron-phonon coupling parameter of the \sp{}-band in dependence on electron temperature and density.
There is an obvious  difference in magnitude to the previous case, as the maximum \sp{}-coupling amounts to only half the \del{}-electron maximum.
For low \sp{}-densities, corresponding to high $\Delta n$ values, the interesting behavior already seen in Fig.~\ref{au_sum} emerges.
At low temperatures, further features in the coupling are indicated for a variety of densities.
However, due to the aforementioned difference in magnitude, they are barely visible in this representation.

Understanding this behavior is more challenging than in the \del{}-band case, as the \sp{}-DOS lacks a similarly defining feature as the large \del{}-band peaks.
To have a better resolution of the density dependence, we examine cuts along fixed temperatures out of the bottom part of Fig.~\ref{au_partial}.

\begin{figure}[tbhp]
	\centering
	\includegraphics[width=0.485\textwidth]{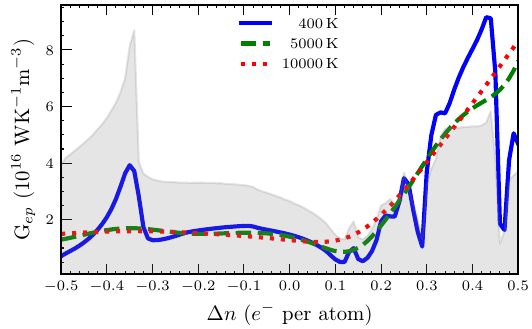}
	\caption{
		Coupling parameter of the \sp{}-electrons at various temperatures in dependence on \sp{}-density and the partial \sp{}-DOS at the chemical potential for 400\,K.
    	There is a clear resemblance between coupling and DOS.}
	\label{sp_mu}
\end{figure}

Figure \ref{sp_mu} displays the electron phonon coupling parameter of the \sp{}-band at three fixed electron temperatures in dependence on density.
The density is represented by \deltan{} defined in Eq.~(\ref{deltan}).
At 400\,K, only the electrons close to the the narrow Fermi edge can noticeable contribute to the electron-phonon coupling.
In that case, the DOS at the chemical potential is of special relevance, as it determines the amount of involved electrons.
To highlight this connection, Fig.~\ref{sp_mu} also shows the \sp{}-DOS at the chemical potential for 400\,K in dependence on the given \deltan{}.
Remember that positive \deltan{} values represent low \sp{}-densities and vice versa.

We find that the behavior of the coupling at 400\,K in Fig.~\ref{sp_mu} closely follows the behavior of the \sp{}-DOS at the chemical potential for the given density.
The features of the DOS are directly reflected in the coupling parameter.
However, this connection is less pronounced for low values of \deltan{}.
This can be explained with the contribution of \del{}-electrons, which participate in the screening of the electron-phonon interaction.
For lower \deltan{}, thus lower \del{}-density, the available phase space of the \del{}-electrons increases, increasing the screening as well.
As can be seen from the matrix element in Eq.~(\ref{PWmatrix}), this also leads to a reduction of the electron-phonon coupling for the \sp{}-electrons.

For the increased electron temperatures of 5000\,K and 10\,000\,K, Fig. \ref{sp_mu} shows that the connection between coupling and DOS is much less direct.
For 5000\,K, the peaks in the coupling appear smoothed out compared to the 400\,K case.
This 5000\,K curve can be understood as the convolution of the low-temperature case with the broadened Fermi edge.

In the highest temperature case, the resulting electron-phonon coupling is completely smooth and lacks any direct reflection of features in the \sp{}-DOS.
The density-dependence of the \sp{}-coupling at such temperatures stems almost exclusively from the screening effect, i.e. the increase of screening for lower \deltan{}.
Note that the chemical potential is a function of both temperature and density and thus the DOS at the chemical potential would look slightly different in the higher temperature cases.

In total, the behavior of \sp{}-coupling could be traced back to a combination of DOS features and screening changes.
The relative importance of these aspects depends heavily on the electron temperature.

\section{Summary and Conclusion}

Overall, we find that the band occupation has a large impact on the electron-phonon coupling of gold.
The increased available phase space volume of \del{}-electrons at lower \del{}-densities leads to a drastic increase in the energy transfer between electrons and phonons.
Characteristic features of the DOS of \sp{}-electrons are reflected in the coupling between \sp{}-electrons and phonons, linked through the density-dependent chemical potential.
Our work shows that the electron temperature alone is not a sufficient predictor of the electron-phonon coupling strength.

\bibliographystyle{ieeetr}
\bibliography{bibfile/all.bib}

\end{document}